\documentclass[10pt, conference, letterpaper]{IEEEtran}
\usepackage{url}
\usepackage{graphicx}
\usepackage{color,xcolor}
\usepackage{clrscode}
\usepackage{multirow}
\usepackage{amsmath}
\usepackage{cases}
\usepackage{subfigure}
\usepackage{caption}
\begin{document}
\newtheorem{DF}{Definition}
\newtheorem{LM}{Lemma}
\newtheorem{THM}{Theorem}
\newtheorem{COR}{Corollary}
\title{Taming Energy Cost of Disk Encryption Software on Data-Intensive Mobile Devices}
\author{
\IEEEauthorblockN{Yang Hu}
\IEEEauthorblockA{
Xi'an Jiaotong University, Xi'an,  China\\
Email: huyang0905@126.com
}
\and
\IEEEauthorblockN{John C.S. Lui}
\IEEEauthorblockA{
The Chinese University of Hong Kong, China\\
Email: cslui@cse.cuhk.edu.hk
}
\and
\IEEEauthorblockN{Wenjun Hu}
\IEEEauthorblockA{
Palo Alto Networks Inc., Singapore\\
Email: mindmac.hu@gmail.com
}
\and
\IEEEauthorblockN{Xiaobo Ma}
\IEEEauthorblockA{
Xi'an Jiaotong University, Xi'an,  China\\
Email: ma.xjtu@qq.com
}
\and
\IEEEauthorblockN{Jianfeng Li}
\IEEEauthorblockA{
Xi'an Jiaotong University, Xi'an,  China\\
Email: jfli.xjtu@gmail.com
}
\and
\IEEEauthorblockN{Xiao Liang}
\IEEEauthorblockA{
Xi'an Jiaotong University, Xi'an,  China\\
Email: qingyuanxingsi@stu.xjtu.edu.cn
}
}
\maketitle
\begin{abstract}
Disk encryption is frequently used to secure confidential data on mobile devices.
However, the high energy cost of disk encryption poses a heavy burden on those devices with limited battery capacity especially when a large amount of data needs to be protected by disk encryption.
To address the challenge, we develop a new kernel-level disk encryption software, \textit{Populus}.
Almost 98\% of \textit{Populus's} encryption/decryption computation is not related with the input plaintext/ciphertext,
so we accomplish the computation in advance during initialization when a consistent power supply is available.
We conduct cryptanalysis on \textit{Populus} and finally conclude that state-of-the-art cryptanalysis techniques fail to break \textit{Populus} in reasonable computational complexity. 
We also conduct energy consumption experiments on \textit{Populus} and dm-crypt, a famous disk encryption software for Android and Linux mobile devices.
The experimental results demonstrate that \textit{Populus} consumes 50\%-70\% less energy than dm-crypt.
\end{abstract}

\section{Introduction}
In recent years, mobile devices, such as smartphones, smartwatches and mobile video surveillance devices~\cite{gualdi2008video}, have become an integral part in our daily life.
Meanwhile, mobile devices are usually facing profound security challenges, especially when being \textit{physically} controlled by attackers.
For example, due to device loss or theft, data leakage in mobile devices happens more frequently than before~\cite{la2013survey}.
To deal with the aforementioned security challenge, mobile devices can encrypt secret data and store its ciphertext locally on itself, which is also known as \textit{disk encryption}~\cite{DiskEncryptionTheory}.
This method attracts extensive attention in industry and academia~\cite{svajcer2014sophos}.
Generally speaking, there are two types of disk encryption solutions: software and hardware solutions. 
This paper mainly focuses on software solutions as they usually have advantages in compatibility and scalability.

However, for data-intensive applications such as mobile video surveillance~\cite{gualdi2008video} and seismic monitor~\cite{zambrano2015distributed}, the whole energy consumption of mobile devices highly rises after applying existing disk encryption software.
One evidence proposed by Li et al. states that for data-intensive applications nearly 1.1-5.9 times more energy is required on commonly-used mobile devices when turning on their disk encryption software~\cite{li2014energy}.
Worse, mobile devices are usually battery-powered in order to improve portability. 
For example, sometimes mobile video surveillance device is equipped on a multi-rotor unmanned aerial vehicle, so battery becomes its sole power supply~\cite{xiao2014video}.
Due to mobile devices' limited battery capacity, existing disk encryption software may strongly affect their normal usage.

The significant energy overhead of existing disk encryption software can be explained by the following two reasons.
First, the ciphers used in existing disk encryption software contain many CPU and RAM operations, which are usually not energy-efficient.
Second, massive data needs to be protected by disk encryption for data-intensive application, which multiplies its energy consumption.
For instance, mobile video surveillance devices need to real-timely record and securely store a large amount of video data~\cite{xiao2014video}.
According to our experiments, nearly 1/3 of energy consumption comes from existing disk encryption software in mobile video surveillance.

In fact, energy consumed by CPU and RAM operations tends to become more prominent than other conventional concerns such as screen and network communication, especially when disk encryption software participates in data-intensive tasks.
About \textit{six years ago},  about 45\%-76\% of daily energy consumption came from screen and GSM when disk encryption software is disabled~\cite{carroll2010analysis}.
However, the distribution of energy consumption has been changed dramatically in recent years due to software\&hardware optimization and usage habit transformation.
A recent study~\cite{xia2015power} shows that for typical usage only about 28\% of energy consumption results from screen and GSM, while CPU and RAM spend about 35\% energy and become the largest energy consumption source in mobile devices when disk encryption is disabled.
In addition, both \cite{carroll2010analysis} and \cite{xia2015power} measures energy consumption without enabling disk encryption function.
So when considering that existing disk encryption software owns many CPU and RAM operations, we believe that the energy consumption percentage of CPU and RAM may be more higher than 35\% if data-intensive mobile devices enable disk encryption software.
Li's experiment results~\cite{li2014energy} exactly verified it.
Hence, to build an energy-efficient mobile system, reducing the energy consumption in disk encryption software is a rational starting point.



To reduce energy consumption of disk encryption software, some researchers try to reduce the number of CPU or RAM operations in disk encryption software.
But it is really challenging to make disk encryption software both energy-saving and cryptographically secure in this way.
Generally speaking, the less computation disk encryption software needs, the less energy it costs, but possibly the more insecure in cryptography.
For example, some trials~\cite{crowley2001mercy} are faced with challenges in terms of cryptography~\cite{fluhrer2002cryptanalysis}.
In existing \textit{cryptographically secure} disk encryption software, the disk encryption software used in Linux and Android, also known as \textit{dm-crypt}, theoretically owns less computation than others.
But our experimental results show that nearly 30\%-50\% of mobile device's energy consumption comes from dm-crypt for typical usage of data collection and transmission.
So the energy consumption of state-of-the-art disk encryption software is still unnegligible.

In this paper, we design and implement a new energy-efficient kernel-level disk encryption software, \textit{Populus}.
The basic idea behind \textit{Populus} is to extract the ``\textit{input-free}" computation from the cipher in disk encryption software and accomplish it during initialization,
where the input-free computation refers to the cipher's operations that are not involved with the input text (i.e., plaintext or ciphertext).
For example, in AES-CBC cipher, its key expansion can be regarded as input-free computation.
The initialization's energy consumption is not considered in this paper
because it is performed only once when a mobile device is first used and a consistent power supply is usually available.
Therefore, the more input-free computation we can extract, the more energy we can save.

However, the ciphers used in existing disk encryption software only have a little input-free computation.
For example, we found that input-free computation of AES-CBC cipher accounts for at most 1\% of all its computation.
To improve the proportion of input-free computation, \textit{Populus} first generates \textit{pseudo random numbers} (\textit{PRNs}) and global matrices in an input-free manner.
Next, it use those PRNs and global matrices to construct temporary matrices and then conduct carefully designed matrix multiplication when encrypting user's privacy.
Each PRN can only participate in disk encryption once so that sufficient PRNs are usually needed for data-intensive application.
To protect those PRNs and matrices, \textit{Populus} encrypts them in an \textit{iterative} manner. 
In this way, \textit{Populus} can save much energy because almost 98\% of its computation is input-free and the residual real-time computation is much smaller than current disk encryption software.
In addition, \textit{Populus} costs acceptable extra storage space (typically $\leq$ 256MB) for those PRNs and matrices.

To assess \textit{Populus} in the respect of cryptographic security and energy efficiency, we conduct cryptanalysis on \textit{Populus} and a series of energy consumption experiments on both \textit{Populus} and dm-crypt.
Finally we find that \textit{Populus} can defend against state-of-the-art cryptanalysis techniques and simultaneously consume less energy than dm-crypt.
Our contribution can be generalized into the following items.
\begin{itemize} 
	\item To the best of our knowledge, this paper is the first work focusing on extracting input-free computation from disk encryption software, which can be used to reduce its energy consumption.
	\item We design and implement an energy-saving kernel-level disk encryption software \textit{Populus} that can both defend against state-of-the-art cryptanalysis techniques and save 50\%-70\% more energy than dm-crypt.
\end{itemize}

The remainder of the paper is organized as follows.
Section~\ref{sec: background} explains why existing disk encryption software is lack of input-free computation and how to improve its proportion in \textit{Populus}.
Section~\ref{sec: details} presents our system \textit{Populus} in detail.
Section~\ref{sec: cryptoanalysis} evaluates the cryptographic security of \textit{Populus}.
Section~\ref{sec:evaluation} presents the experimental results of mobile devices' energy consumption.
Section~\ref{sec: related work} summarizes related work.
Concluding remarks then follow.

\section{\textbf{Input-Free Computation}}
\label{sec: background}
In this section, we present more details about input-free computation.
We first introduce the design consideration of existing disk encryption software (e.g., dm-crypt) and explain why they are lack of input-free computation.
Then, we show the basic idea of \textit{Populus} and illustrate why it can improve the proportion of input-free computation.

Existing disk encryption software including dm-crypt is usually based on \textit{tweakable scheme}~\cite{Shai:Halevi2003A}, where each disk sector should correspond to an independent key used only for its encryption and decryption.
However, in practice, a user only provides one master key.
To solve this problem, most of existing disk encryption software achieves tweakable scheme in the following two steps: \textbf{1)} produce sector-specific keys based on master key and sector ID; \textbf{2)} use sector-specific key to encrypt certain disk sector with a block cipher.

Due to the fact that attackers can get multiple (plaintext, ciphertext) pairs in the same disk sector, they can conduct \textit{chosen-plaintext attack} (\textit{CPA})~\cite{ChosenPlaintextAttack} by exploiting (plaintext, ciphertext) pairs sharing same sector-specific key.
Hence, to secure the whole crypto system, the block cipher in \textbf{2)} must have the ability to defend against CPA.
To achieve this, one effective solution is to construct a block cipher in \textit{substitution-permutation network} (\textit{SPN})~\cite{william2011cryptography}. 
Unfortunately, we find that nearly all SPN-based block ciphers have a little input-free computation because their core components, \textit{substitution box} and \textit{permutation box}, directly or indirectly rely on input.

To improve input-free computation, we give up aforementioned tweakable scheme and SPN when designing \textit{Populus}.
Instead, we construct \textit{Populus} based on \textit{nonce-based scheme}~\cite{rogaway2004nonce}.
\textit{Populus's} core design can be briefly described as follows: \textbf{a)} for $i^{th}$ encryption, produce an independent temporary key based on master key and $i$; \textbf{b)} use the temporary key and a light-weight block cipher to accomplish $i^{th}$ encryption.
Our design has four advantages.
First, it is compatible to tweakable scheme.
Second, it makes attackers hard to get multiple (plaintext,ciphertext) pairs sharing same key, and thereafter basically eliminate the threat from CPA.
Third, nearly all procedures in \textbf{a)} are input-free.
Forth, SPN becomes unnecessary in \textbf{b)}, which gives us more freedom to design a light-weight block cipher owning much input-free computation.
As a trade-off, our scheme needs extra storage space for input-free computation.
Fortunately, the storage space can be reduced to an acceptable level by carefully designing the temporary key production method in \textbf{a)} and the block cipher in \textbf{b)}.
In Section~\ref{sec: details}, we complement details regarding the design and implementation of \textit{Populus}.
\section{\textbf{Populus: An Energy-Saving Disk Encryption Software System}}\label{sec: details}
\begin{figure}
\centering
\includegraphics[width=3.4in,height=2.6in]{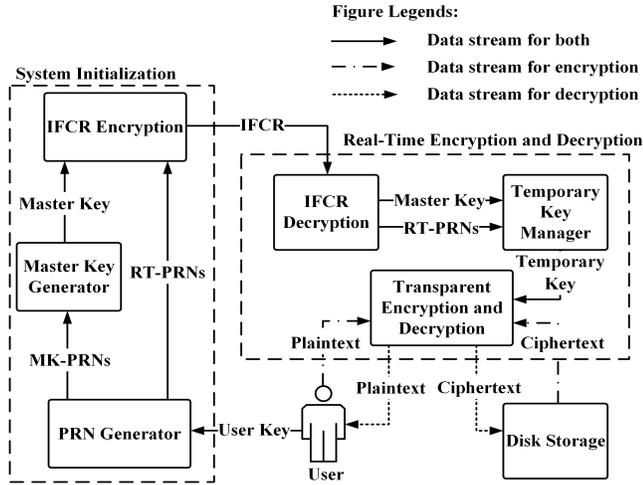}
\caption{Overview for \textit{Populus}}
\label{fig:EPE}
\end{figure}

The overview of \textit{Populus} is shown in Fig.~\ref{fig:EPE}.
\textit{Populus} consists of two parts: \emph{system initialization} and \emph{real-time encryption/decryption}.
We perform system initialization once when a mobile device is first used and we assume that a consistent power supply is available and the energy consumption is not a concern during system initialization.
\textit{Populus} initially accomplishes all input-free computation and stores its result on disk,
which is used for processing real-time encryption and decryption requests later.
\textit{Populus} works at a \textit{512-byte disk sector} level,
and it allows users to manually configure \textit{private disk sectors},
which store users' confidential information.
For each private disk sector,
\textit{Populus} initially assigns it a \textit{temporary key}, which is used for encrypting/decrypting the confidential data on it.
Each temporary key can only be used for one encryption.
If a sector has 'consumed' its temporary key due to encryption, \textit{Populus} will recycle its current temporary key and allocate a new temporary key for its next encryption.
We design \textit{Populus} for 64-bit systems because 64-bit processors are popular for mobile devices~\cite{Android64bit}.
Throughout the paper, the default value of a number's size is 64 bits unless stated otherwise.
For the ease of reading, we list notations used throughout the paper in Table~\ref{table: symbols for populus}.
Next, we introduce each part of \textit{Populus} in detail.
\begin{table}
	\centering
	\caption{Table of notations.}
	\begin{tabular}{|l|l|} \hline
		\textbf{Notation}&\textbf{Meaning}\\ \hline
		\multirow{3}{*}{$P$ or $P_{i}$}
		& A 512-byte plaintext consisting of $64$ numbers. \\
		& In particular, $i$ is used to differentiate multiple\\
		& plaintexts. \\ \hline
		\multirow{3}{*}{$P^{(j)}$ or $P^{(j)}_{i}$}
		& The $j^{th}$ number in $P$ or $P_{i}$.\\
		& In particular, $i$ is used to differentiate multiple\\
		& plaintexts. \\ \hline
		\multirow{3}{*}{$C$ or $C_{i}$}
		& A 512-byte ciphertext consisting of $64$ numbers.\\
		& In particular, $i$ is used to differentiate multiple\\
		& ciphertexts. \\ \hline
		\multirow{3}{*}{$C^{(j)}$ or $C^{(j)}_{i}$}
		& The $j^{th}$ number of $C$ or $C_{i}$.\\
		& In particular, $i$ is used to differentiate multiple \\
		& ciphertexts. \\ \hline
		\multirow{1}{*}{$U$}
		& Master key that consists of 125 $2\times 2$ matrices.\\\hline
		\multirow{1}{*}{$U^{(i)}$}
		& The $i^{th}$ matrix in $U$.\\\hline
		\multirow{3}{*}{$M$ or $M_{i}$}
		& A temporary key that consists of 125 $2\times 2$ matrices.\\
		& In particular, $i$ is used to differentiate multiple \\ & temporary keys. \\ \hline
		\multirow{3}{*}{$M^{(j)}$ or $M^{(j)}_{i}$}
		& The $j^{th}$ matrix in $M$ or $M_{i}$.\\
		& In particular, $i$ is used to differentiate multiple\\
		& temporary keys. \\ \hline
		\multirow{1}{*}{$R$}
		& RT-PRNs\\ \hline
		\multirow{1}{*}{$R^{(i)}$}
		& The $i^{th}$ PRN in $R$.\\ \hline
		\multirow{1}{*}{$E(P,M)$}
		& Encryption for one plaintext $P$.\\ \hline 
		\multirow{1}{*}{$E(P_{i},M_{i})$}
		& Encryption for one plaintext $P_{i}$. \\ \hline
		\multirow{1}{*}{$E(P_{1:\theta},M_{1:\theta})$}
		& $(E(P_{1},M_{1}),\ldots,E(P_{\theta},M_{\theta}))$\\ \hline
		\multirow{1}{*}{$D(C,M)$}
		& Decryption for one ciphertext $C$.\\ \hline 
		\multirow{1}{*}{$D(C_{i},M_{i})$}
		& Decryption for one ciphertext $C_{i}$.\\ \hline
		\multirow{1}{*}{$D(C_{1:\theta},M_{1:\theta})$}
		& $(D(C_{1},M_{1}),\ldots,D(C_{\theta},M_{\theta}))$\\ \hline
		\multirow{2}{*}{$n$}
		& The numbers of disk sectors storing input-free\\ & computation result (IFCR). \\ \hline
		\multirow{1}{*}{$S_{i}$}
		& The set containing all $i$-byte sequences.\\ \hline
	\end{tabular}
	\label{table: symbols for populus}
\end{table}
\subsection{\textbf{System Initialization}}
The system initialization includes three input-free modules: \textit{PRN generator}, \textit{master key generator} and \textit{IFCR encryption}.
Here, \textit{IFCR} is the abbreviation of \textit{input-free computation result}.
PRN generator produces PRNs,
which are basic for generating master key and real-time encryption/decryption.
Next, \textit{Populus} encrypts IFCR and then stores it on disk.
\subsubsection{\textbf{PRN Generator}}
To produce PRNs, we use Salsa20/12 stream cipher,
which has been extensively studied and found to produce PRNs of very high quality~\cite{bernstein2008salsa20}.
Salsa20/12 requires a 320-bit input,
hence we use the SHA3 algorithm~\cite{SHA3} to map a user's arbitrary-length key into a 384-bit number, and extract the first 320-bit \textit{hash key} as the input of Salsa20/12 stream cipher.
PRNs are mainly used for master key production and real-time encryption/decryption, which are separately named \textit{MK-PRNs} and \textit{RT-PRNs}.
\subsubsection{\textbf{Master Key Generator}}
\textit{Populus} generates master key using MK-PRNs.
We define a square matrix $A$ is $2^{64}$ \textit{modular invertible} when there exists a matrix $B$ such that $AB = I \mod 2^{64}$,
where $I$ is the identity matrix.
If this is the case,
then the matrix $B$ is uniquely determined by $A$ and is called the modular inverse of $A$ (mod $2^{64}$).
For simplicity, we denote it by $A^{-1}$ in this paper.
We denote master key as $U=(U^{(1)}, \ldots, U^{(125)})$,
where each $U^{(i)}=\begin{pmatrix}
u_{1,1}^{(i)} & u_{1,2}^{(i)}\\
u_{2,1}^{(i)} & u_{2,2}^{(i)}
\end{pmatrix}$, $1\le i\le 125$, is a $2\times 2$ matrix and $U^{(i)}$ is \textit{modular invertible},
which is critical for the real-time encryption and decryption discussed later.

We randomly select matrices $U^{(1)}, \ldots, U^{(125)}$ from the set of \textit{modular involutory} matrices based on the Acharya's method~\cite{acharya2009involutory}.
Here a modular involutory matrix is defined as a matrix whose modular invertible matrix is itself.
Since there exists $7.66\times 10^{38}$ modular involutory matrices~\cite{overbey2005keyspace}, the number of all possible $U$ is $(7.66\times 10^{38})^{125}\approx 3.38\times 10^{4860}$,
which is much larger than the size of our hash key space (i.e., $2^{320}\approx 2.14\times 10^{96}$).
Therefore, the master key is more difficult to brutally crack than the hash key.
\subsubsection{\textbf{IFCR Encryption and Decryption}}\label{sec: IFCRE}
To protect IFCR (i.e., RT-PRNs and master key), \textit{Populus} encrypts them and then stores them on disk.
During real-time encryption/decryption, \textit{Populus} decrypts master key and RT-PRNs from disk.
Later, we will introduce more detail in Section~\ref{sec: IFCRD}.
\subsection{\textbf{Real-Time Encryption and Decryption}}
\textit{Populus} performs disk encryption/decryption when the file system writes/reads data on disk in real time.
We introduce each of its modules as follows:
\subsubsection{\textbf{Transparent Encryption and Decryption}}\label{sec: XM_MMM}
Our transparent encryption and decryption is based on matrix multiplication in \textit{modular linear algebra}~\cite{eisenberg1999hill}.
In cryptography, matrix multiplication has achieved Shannon's diffusion~\cite{shannon1949communication} and it dissipates statistical structure of the plaintext into long-range statistics of the ciphertext to thwart cryptanalysis based on statistical analysis~\cite{william2011cryptography}.
However, matrix multiplication is usually computationally intensive.
For example, a general matrix multiplication between a $64\times 64$ matrix and a $64\times 1$ matrix requires $64\times 64+64\times 63+128=8256$ operations.

To reduce its computation, \textit{Populus} only constructs $125$ $64\times 64$ sparse matrices $H^{(i)}=\begin{pmatrix}
I_{62-|63-i|} & \textbf{0} & \textbf{0} \\
\textbf{0} & M^{(i)} & \textbf{0} \\
\textbf{0} & \textbf{0} & I_{|63-i|}
\end{pmatrix}$ where $i\in\{1,\ldots,125\}$, $I_{i}$ is the $i$-dimensional identity matrix, and $M^{(i)}$ is a $2\times 2$ modular invertible matrix.
Then \textit{Populus} computes $H^{(125)}\ldots H^{(1)}P$ as encryption or $(H^{(1)})^{-1}\ldots (H^{(125)})^{-1}C$ as decryption where $P$ is a $64\times 1$ matrix as one 512-byte plaintext and $C$ is a $64\times 1$ matrix as one 512-byte ciphertext.
Exploiting $H^{(i)}$ is a sparse matrix, $125$ $64$-dimensional matrix multiplications can be simplified to $125$ $2$-dimensional matrix multiplications.
The simplified encryption and decryption only consists of $125\times (2\times 2 + 2\times 1)+128=868$ operations.

\begin{figure*}
    \centering
    \includegraphics[height=6.4in, width=7.6in]{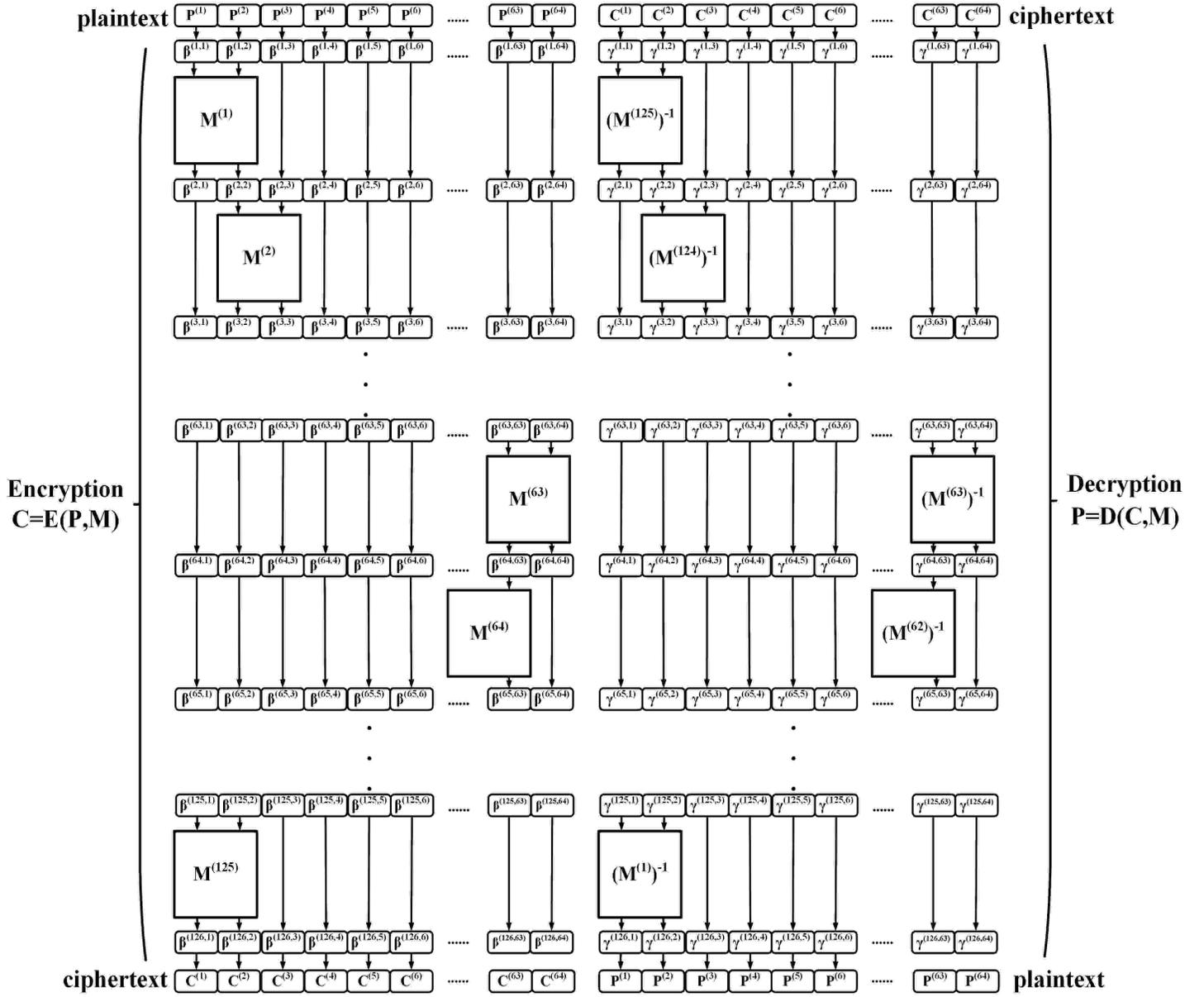}
    \caption{An illustration of transparent encryption and decryption}
    \label{fig:mm_se}
\end{figure*}

Next, we describe our transparent encryption and decryption in more detail.
Fig.~\ref{fig:mm_se} presents its full view.
Let $P=(P^{(1)},...,P^{(64)})^{T}$, $C=(C^{(1)},...,C^{(64)})^{T}$, and $M=(M^{(1)},\ldots,M^{(125)})$ denote its plaintext, ciphertext, and temporary key respectively,
where $P^{(i)}$ is the $i^{th}$ number in the plaintext, $C^{(i)}$ is the $i^{th}$ number in the ciphertext and $M^{(i)}=\begin{pmatrix}
m^{(i)}_{1,1} & m^{(i)}_{1,2}\\
m^{(i)}_{2,1} & m^{(i)}_{2,2}
\end{pmatrix}$ is the $i^{th}$ $2\times 2$ matrix in $M$.
For simplicity, we use the notation $[m]_{n}$ to denote the function $m \mod n$, i.e., $[m]_{n} = m \mod n$.
The encryption function $E(P,M)$ works as follows:
We first set $\beta^{(1,j)}=P^{(j)}$ and then iteratively compute $\beta^{(i+1,j)}$, $1\le i\le 125$, $1\le j\le 64$, as
\begin{eqnarray}
\notag
\beta^{(i+1,j)}=
\begin{cases}
[\beta^{(i,j)} m_{1,1}^{(i)} + \beta^{(i,j+1)}  m_{1,2}^{(i)}]_{2^{64}}, &j=i, 126-i\cr
[\beta^{(i,j-1)} m_{2,1}^{(i)} + \beta^{(i,j)} m_{2,2}^{(i)}]_{2^{64}}, &j=i+1, 127-i\cr
\beta_{i,j}, &\text{otherwise.}
\end{cases}
\end{eqnarray}
Finally, we set $E(P,M)=(\beta^{(126,1)}$,\ldots, $\beta^{(126,64)})^{T}$.

The decryption $D(C,M)$ function works as follows:
Let $(M^{(i)})^{-1}=\begin{pmatrix}
l_{1,1}^{(i)} & l_{1,2}^{(i)}\\
l_{2,1}^{(i)} & l_{2,2}^{(i)}
\end{pmatrix}$ and $k=126-i$.
We set $\gamma^{(1,j)}=C^{(j)}$ and then iteratively compute $\gamma^{(i+1,j)}$, $1\le i\le 125$, $1\le j\le 64$, as
\begin{eqnarray}
\notag
\gamma^{(i+1,j)}=
\begin{cases}
[\gamma^{(i,j)}l_{1,1}^{(k)} + \gamma^{(i,j+1)}l_{1,2}^{(k)}]_{2^{64}}, &j=i, 126-i\cr
[\gamma^{(i,j-1)} l_{2,1}^{(k)} + \gamma^{(i,j)} l_{2,2}^{(k)}]_{2^{64}},&j=i+1, 127-i\cr
\gamma^{(i,j)}, &\text{otherwise.}
\end{cases}
\end{eqnarray}
Finally, we set $D(C,M)=(\gamma^{(126,1)}, \ldots, \gamma^{(126,64)})^{T}$.
\subsubsection{\textbf{Temporary Key Manager}}
Each temporary key consists of $125\times 2\times 2\times 8 = 4000$ bytes,
therefore storing all temporary keys requires a large storage space.
To solve this problem, \textit{Populus} computes its $M$ based on $U$, $R_{2j-1}$ and $R_{2j}$ for $j^{th}$ encryption, where $R=(R_{1},...,R_{d})$ denote RT-PRNs, $R_{i}, 1\le i\le d$, is a pseudo random number, and $d$ is the size of $R$.
Note that $U$ is shared by all temporary keys' construction and its size is $4000$ bytes,
so on average, we only need about $16$ bytes for storing a temporary key.
Then, for $j^{th}$ encryption, we compute each $m_{p,q}^{(i)}$ ($p,q=\{1,2\}$) in $M^{(i)}$, as
\begin{eqnarray}
\label{equ:mg}
m_{p,q}^{(i)}=
\begin{cases}
[2(u_{p,q}^{(i)}\oplus R_{2j-1})+[u_{p,q}^{(i)}]_{2}]_{2^{64}} ,& i=1,\cr
[2(u_{p,q}^{(i)}\oplus R_{2j-1}\oplus R_{2j})+[u_{p,q}^{(i)}]_{2}]_{2^{64}} ,& i=63,\cr
[2(u_{p,q}^{(i)}\oplus R_{2j})+[u_{p,q}^{(i)}]_{2}]_{2^{64}} ,& i=125,\cr
u_{p,q}^{(i)}, & \text{otherwise}. \cr
\end{cases}    
\end{eqnarray}
\begin{THM}
	$M^{(i)}$ is modular invertible.
	\begin{IEEEproof}
From~\cite{eisenberg1999hill}, we find that $M^{(i)}$ is modular invertible if and only if $|M^{(i)}|$ and $2^{64}$ are co-prime, where $|M^{(i)}|$ denotes the determinant of matrix $M^{(i)}$.
Therefore, $M^{(i)}$ is modular invertible when $|M^{(i)}|$ is an odd number.
Next, we prove $|M^{(i)}|$ is an odd number.
From Eq.~(\ref{equ:mg}), we can easily find that $m_{p,q}^{(i)}$ and $u_{p,q}^{(i)}$ have the same parity for any $p,q\in \{1,2\}$.
Thus, $|M^{(i)}|=m_{1,1}^{(i)}m_{2,2}^{(i)} - m_{1,2}^{(i)}m_{2,1}^{(i)}$ and $|U^{(i)}|=u_{1,1}^{(i)}u_{2,2}^{(i)} - u_{1,2}^{(i)}u_{2,1}^{(i)}$ have the same parity.
$U^{(i)}$ is modular invertible, so we know $|U^{(i)}|$ and $2^{64}$ are co-prime from~\cite{eisenberg1999hill}.
Thus, $|U^{(i)}|$ and $|M^{(i)}|$ are both odd numbers.
\end{IEEEproof}
\end{THM}

Considering that RT-PRNs can only be used once, $d$ should be as large as possible in order to securely store mass data.
But if $d$ is too large, RT-PRNs will occupy a lot of storage space so that there may be no enough space for user's data.
To mitigate this contradiction, \textit{Populus} only stores a balanced amount of RT-PRNs that can support real-time encryption/decryption before battery uses up and then replenishes RT-PRNs during device charging or battery replacement.

After applying our method, only small storage space of RT-PRNs is able to satisfy most of applications in practice.
Suppose that on average mobile devices require to securely store $l$-byte data each day and can work $t$ days without enabling disk encryption. 
\textit{Populus} needs at most $d=lt/256$ pseudo random numbers in RT-PRNs.
For example, as for smartphone, we let $t=4$ and $l=2^{31}$ so that only $256$ MB are required to store $d=2^{25}$ pseudo random numbers.
\subsection{\textbf{Iterative Encryption and Decryption on IFCR}}\label{sec: IFCRD}
In Section~\ref{sec: IFCRE}, we have briefly introduced the function of IFCR encryption and decryption.
However, IFCR decryption may cost much energy if we choose existing block ciphers as its encryption/decryption algorithm. 
For example, if \textit{Populus} uses AES-CBC to encrypt IFCR, nearly all encrypted IFCR should be decrypted for each time of transparent encryption, which obviously costs lots of energy.

To reduce the aforementioned energy cost, we propose a dedicated encryption method called \textit{iterative encryption} for IFCR protection.
The basis idea of iterative encryption comes from our observation that \textit{Populus} only needs one new master key ($4000$ bytes) and $k$ new RT-PRNs ($16k$ bytes) when encrypting $k$ disk sectors (512 bytes for each disk sector) whose data is never changed.
Considering that master key and RT-PRNs are never changed once generated, we iteratively encrypt them as follows:
\textbf{a)} If IFCR only occupies $k\leq 9$ disk sectors in all, \textit{Populus} directly encrypts them through a SPN-based cipher (e.g., AES-CBC); \textbf{b)} If IFCR occupies $k>9$ disk sectors, \textit{Populus} produces another new IFCR including one new master key and $\lceil(16k+4000)/512\rceil$ new RT-PRNs and use them to encrypt original IFCR through our proposed transparent encryption method; \textbf{c)} Encrypt new IFCR by repeating \textbf{a)} and \textbf{b)}.
As for \textit{iteration decryption}, just reverse the whole process of iteration encryption.

We can use \textit{master method} to prove that the computation complexity of our iterative encryption/decryption is $O(log(n))$.
Here, $n$ denotes the number of disk sectors occupied by IFCR.
Compared with AES-CBC which needs $O(n)$ computation in same task, our iterative encryption/decryption save much energy. 

\section{Security Analysis of Populus}\label{sec: cryptoanalysis}
To rigorously assess \textit{Populus's} security, we first introduce widely-used security definitions in cryptanalysis theory such as \textit{message indistinguishability}.
Based on those definitions, we analyze whether \textit{Populus} can effectively defend against state-of-the-art cryptanalysis techniques such as \textit{linear}~\cite{matsui1994linear}, \textit{differential}~\cite{biham1991differential}, \textit{algebra}~\cite{ferguson2001simple}, \textit{slide}~\cite{biryukov1999slide}, and \textit{Biclique attacks}~\cite{khovratovich2012bicliques} .
The analysis results show that all those techniques fail to break \textit{Populus} in reasonable computational complexity.
\subsection{\textbf{Security Definition}}
\label{sec: sd}
Our security analysis mainly focuses on message indistinguishability, an important property in cryptography that most of existing security analysis works always discuss.
Message indistinguishability of \textit{Populus} can be briefly explained as the difficulty to distinguish two groups of ciphertexts.
In detail, we assume that \textit{Populus} has produced a group of temporary keys denoted by $M_{1:\theta}=(M_{1},\ldots,M_{\theta})$ with the same hashed key randomly chosen by a user and then encrypted two groups of 512-byte plaintexts denoted by $P_{1:\theta}=(P_{1},\ldots,P_{\theta})$, $P_{1:\theta}^{'}=(P^{'}_{1},\ldots,P^{'}_{\theta})$ with $M_{1:\theta}$ and get $C_{1:\theta}=(E(P_{1},M_{1}),\ldots,E(P_{\theta},M_{\theta}))$ and $C_{1:\theta}^{'}=(E(P^{'}_{1},M_{1}),\ldots,E(P^{'}_{\theta},M_{\theta}))$).
Provided with $C_{1:\theta}$ and $C_{1:\theta}^{'}$, $P_{1:\theta}$, $P_{1:\theta}^{'}$, and \textit{Populus's} encryption algorithm, attackers try to design a distinguisher that can propose a correct correspondence between $P_{1:\theta}$,$P^{'}_{1:\theta}$ and $C_{1:\theta}$ and $C_{1:\theta}^{'}$.
Then we informally conclude that \textit{Populus} is message indistinguishable if attackers can't accomplish the distinguishing work in both low computational complexity and high success probability without any prior knowledge of $M_{1:\theta}$ and the hashed key.
 
Next, we give a formal definition of \textit{Populus's} message indistinguishability.
\begin{DF}
\label{DF: SECURE}
\textit{Populus} is $(t,\epsilon,\theta)$ message indistinguishable against an attack method $Adv$ defined as $\{sequences~of~512~bytes\}^{\theta}\rightarrow \{0,1\}$ if and only if the computational complexity of $Adv$ is not more than $t$ and for every $P_{1:\theta}$, $P_{1:\theta}^{'}$, and $C_{1:\theta}$, $C_{1:\theta}^{'}$,
\begin{equation}
|\mathcal{P}(Adv(C_{1:\theta})=1)- \mathcal{P}(Adv(C_{1:\theta}^{'})=1)|\leq \epsilon.
\end{equation}
\end{DF}
In Def.~\ref{DF: SECURE}, the values of $t$, $\epsilon$ and $\theta$ are strongly linked to the real-world security of \textit{Populus}.
From Luca's suggestion~\cite{trevisan2009lecture}, we can get that typical parameters adopted in practical secure crypto system follows $t\leq 2^{80}$, $\epsilon\leq 2^{-60}$ and $\theta\leq t$.
We will later discuss \textit{Populus's} security against certain attack method based on Def.~\ref{DF: SECURE} and Luca's suggestion.

In addition, when discussing \textit{Populus's} security, we assume that our pseudo random number generator (i.e., Salsa20/12) is secure so we don't conduct secure analysis on it.
We don't discuss attack techniques out of cryptography such as DMA-based attack, cold boot attack and evil maid attack because they are beyond this paper's scope.
\subsection{\textbf{Linear Attack and Differential Attack}}
Linear attack~\cite{matsui1994linear} and differential attack~\cite{biham1991differential} are two powerful cryptanalysis techniques towards block ciphers.
Both of the two techniques are chosen-plaintext attacks exploiting the design defects in \textit{S-box}.
Here, S-box is a widely-used component in block ciphers that substitutes its input bit sequence with another bit sequence in same length as its output.
For example, a function $f(x)=(x+1)mod(256),x\in\{0,\ldots,255\}$ is a S-box.

Even though linear and differential attacks have broken various kinds of block ciphers such as DES~\cite{matsui1994first,biham2012differential}, it is hard for them to distinguish messages protected by \textit{Populus} in reasonable computational complexity in the eye of our security definitions in Section~\ref{sec: sd}.
In \textit{Populus}, each S-box is hidden from attackers and closely randomly chosen from a huge S-box space containing at least $2^{1600}$ S-boxes, which makes it computationally impractical to analyze all possible S-boxes. 
Circumventing the useless brute force, some attackers may endeavor to collect special (plaintext,ciphertext) pairs whose corresponding S-boxes are the same and then consider linear or differential characteristics.
However, they still requires to distinguish ciphertexts encrypted by different S-boxes before exploiting linear or differential characteristics.
Obviously, it is an infinite logic loop.
Hence, we conclude that linear and differential attacks can't effectively break \textit{Populus}.

Note that \textit{zero correlation linear attack}~\cite{bogdanov2012zero}, \textit{Boomerang attack}~\cite{Wagner1999The}, \textit{impossible differential attack}~\cite{lu2008new}, \textit{higher-order differential attack}~\cite{Duan2011Higher}, \textit{truncated differential attack}~\cite{Lars1994Truncated} and \textit{differential-linear attack}~\cite{Biham2002Enhancing} are derivative from linear attack or differential attack.
After finding those attack methods still can't elegantly deal with \textit{Populus's} multiple S-Boxes, we conclude that those existing derivatives of linear and differential attacks can't break \textit{Populus} in an effective manner.

\subsection{\textbf{Algebra Attack}}
Algebra attack pays close attention on the algebraic system adopted by a cipher and then break the cipher by exploiting its algebraic characteristics~\cite{ferguson2001simple}.
In practice, existing ciphers on linear system are easier to break because linear system has been fully studied.
For example, Hill cipher, a classical block cipher based on modular linear algebra, is not secure against algebra-based chosen-plaintext attack through easily solved linear transformation.
Considering that \textit{Populus} is also constructed on linear system, algebra attack seems to imperil \textit{Populus} more intensely than other attack methods.

However, \textit{Populus's} algebraic system is a volatile linear system, which substantially reduces its conspicuousness of linear characteristics.
We propose a thorough cryptanalysis based on linear-based algebra attack.
The cryptanalysis results show that \textit{Populus} can't be broken by existing linear-based algebra attack methods in reasonable computational complexity.
Its whole process is shown in Appendix~\ref{APPENDEX: Algebra}. 
\subsection{\textbf{Other Attacks}}
Slide attack is another excellent cryptanalysis technique and can only be used to analyze the ciphers that constitute multiple rounds and each round shares same key~\cite{biryukov1999slide}.
For example, DES, who consists of 16 rounds and the keys of every rounds remain equal, can be broken by slide attack~\cite{biryukov1999slide}.
Given that \textit{Populus} consists of $125$ matrix multiplications which can be regarded as $125$ so-called rounds, it seems that slide attack may break \textit{Populus}. 
However, in \textit{Populus}, the probability of the equivalence among 'keys' (i.e., $M^{(i)},i\in\{1,\ldots,125\}$) in all rounds is lower than $2^{-36000}$.
Given the precondition of slide attack is nearly impossible, we don't think that slide attack is suitable to break \textit{Populus}.

Biclique attack is a distinguished chosen-plaintext attack that can theoretically attack full AES-128 with the computational complexity $2^{126.1}$~\cite{bogdanov2011biclique}.
It can also attack \textit{Populus} by skillfully searching correct $M_{1:\theta}$ in the \textit{meet-in-the-middle} strategy~\cite{WikiMITM}.
However, all $M_{i}^{(j)}$ are nearly independent and randomly chosen, which extremely extends its search space (i.e., at least $2^{4097}$).
Due to the unacceptable search space, we conclude that Biclique attack can't break \textit{Populus} in rational computational complexity.

In conclusion, we have studied five popular cryptanalysis techniques and find that \textit{Populus} successfully defends against them.
We don't discuss other existing cryptanalysis techniques, for they are not quite matched to \textit{Populus}.
\section{Energy Consumption Evaluation}\label{sec:evaluation}
In this section, we use \textit{Monsoon power monitor}~\cite{monsoon} to measure energy consumption of the whole mobile device and estimate the energy cost by disk encryption software.
We choose Google Nexus 4 smartphone with Android 5.0 OS as our tested mobile device.
To compared with Populus proposed in this paper, dm-crypt is chosen as the baseline for the following two reasons.
First, the architecture of dm-crypt is similar to other popular disk encryption software and their computation is close.
So we can use dm-crypt as a representative of existing disk encryption software. 
Second, dm-crypt is compatible with Android.
So it is convenient for us to conduct energy consumption experiments on our Google Nexus 4 smartphone.
\subsection{Evaluation on Typical Usages for Mobile Device}
\label{sec: ec whole device}
We conduct a series of experiments to measure the energy consumption of mobile device's typical usage.
Through those experiments, we can verify whether enabling dm-crypt tremendously raises the whole device's energy consumption and whether \textit{Populus} can mitigate it.

We choose Google Nexus 4 smartphone with Android 5.0 OS as our tested mobile device. 
We also design three configurations for the mobile device: only enabling dm-crpyt, only enabling \textit{Populus} and disabling any disk encryption.
For each configuration, we measure the mobile device's whole energy consumption in four typical usage: video recording, video playing, data sending throgh WIFI, data receiving through WIFI.
As for video playing and recording, video format is \textit{mp4}, video resolution is 480$\times$270, the choices of video length are 50min, 100min, 150min and 200min and video quality is of high definition.
As for WIFI network, the choices of transferred data size are 256MB, 512MB, 768MB,$\ldots$, 2048MB.

\begin{figure}
\centering
\includegraphics[height=2.2in, width=3.4in]{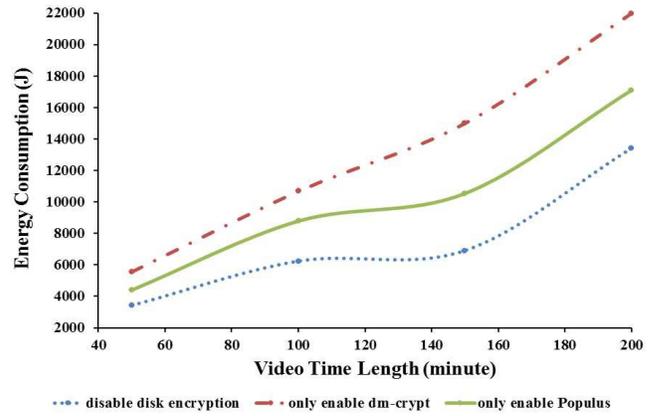}
\caption{Energy consumption of video playing}
\label{fig:videoplay}
\end{figure}

Then we introduce our experiments separately.
Video playing is a common function for handheld mobile device such as smartphone and its energy consumption status is shown in Fig.~\ref{fig:videoplay}.
Note that when playing an encrypted video, decryption is necessary so that part of energy consumption comes from \textit{Populus} or dm-crypt if they are enabled.
As you can see, nearly 1/2 of energy is cost by dm-crypt and \textit{Populus} can reduce it to nearly 1/4.

\begin{figure}
\centering
\includegraphics[height=2.3in, width=3.4in]{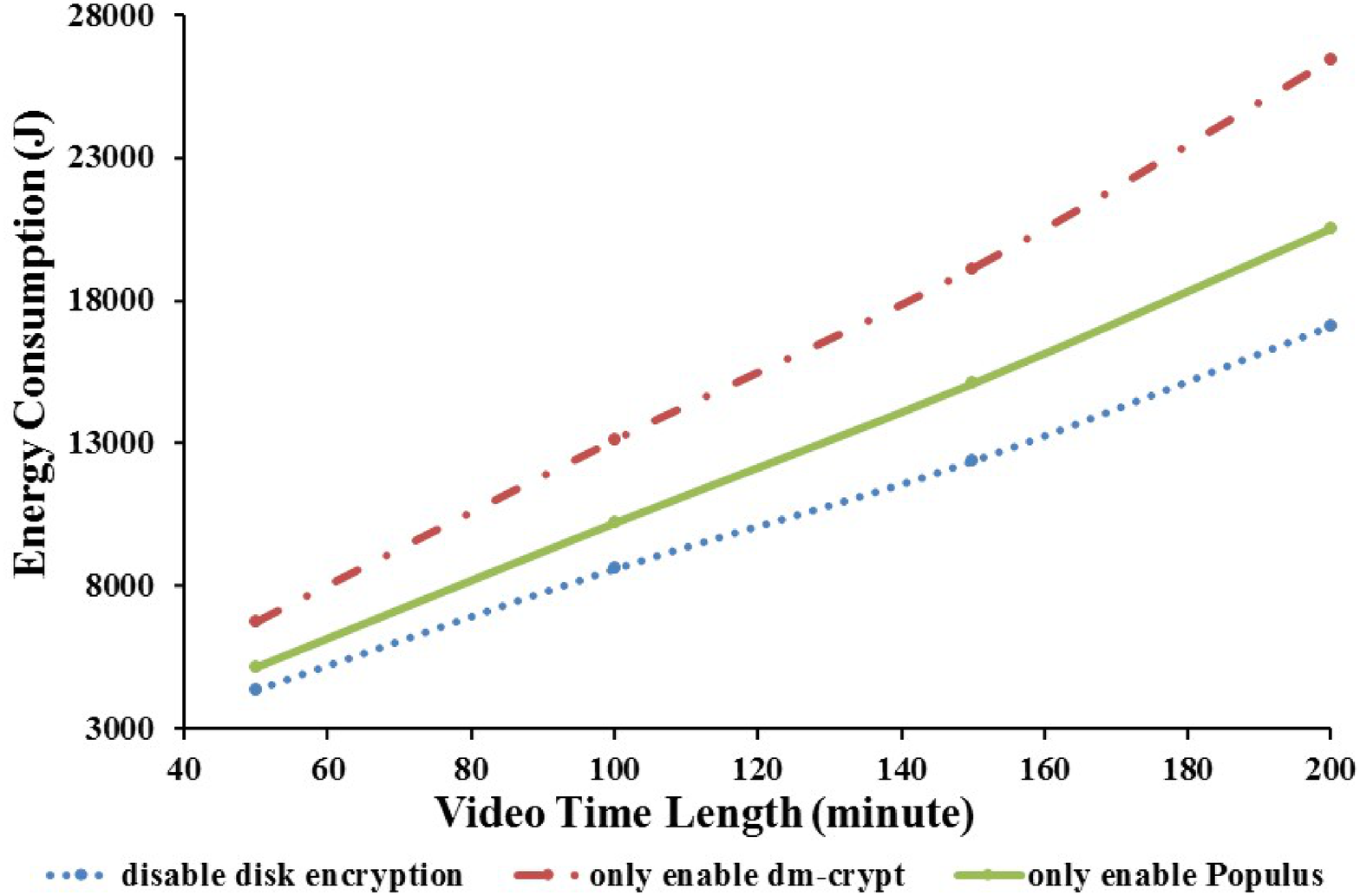}
\caption{Energy consumption of video recording}
\label{fig:videorecord}
\end{figure}

We also present relevant experimental results of video recording in Fig.~\ref{fig:videorecord}, as video recording on mobile device is widely used in personal life, industry and military (e.g., mobile video surveillance~\cite{gualdi2008video}).
Obviously when recording a secret video, disk encryption is necessary so that part of energy consumption comes from \textit{Populus} or dm-crypt if they are enabled.
Our experimental results show that nearly 1/3 of energy is cost by dm-crypt and \textit{Populus} can reduce it to nearly 1/6.

\begin{figure}
\centering
\includegraphics[height=2.3in, width=3.4in]{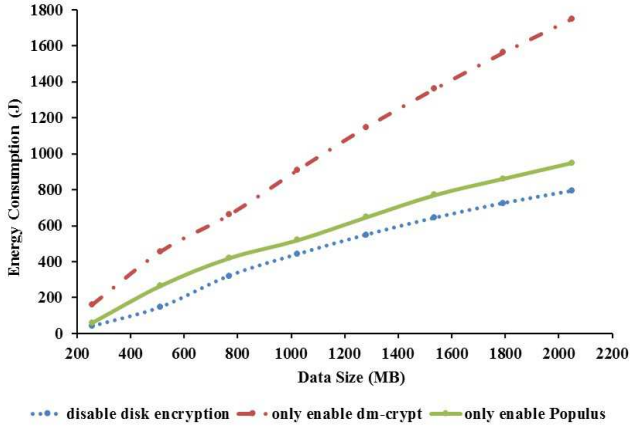}
\caption{Energy consumption of data sending through WIFI}
\label{fig:networksend}
\end{figure}

As for network data transference, Fig.~\ref{fig:networksend} demonstrates mobile device's energy consumption when it sends data to remote terminal through WIFI network.
Here, data has been encrypted by disk encryption software in advance so that data decryption before network transference should be considered if disk encryption software is enabled.
Apparently, there is an approximate linear relation between transferred data size and mobile device's energy consumption.
On average, 51\% of energy consumption on mobile device is cost by dm-crypt and \textit{Populus} can reduce it to 20\%.

\begin{figure}
\centering
\includegraphics[height=2.1in, width=3.4in]{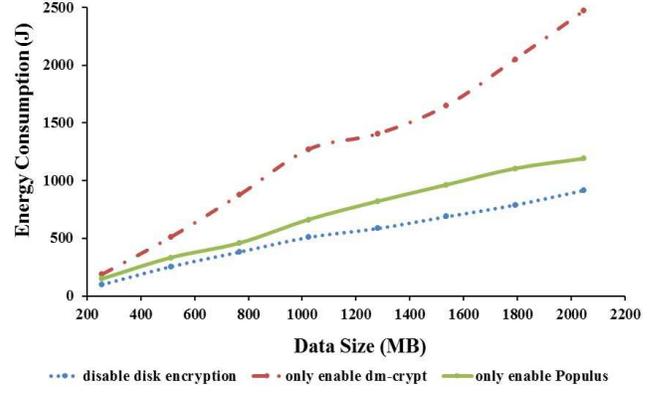}
\caption{Energy consumption of data receiving through WIFI}
\label{fig:networkreceive}
\end{figure}

Fig.~\ref{fig:networkreceive} shows mobile device's energy consumption when it receives data from remote terminal through WIFI network.
Here, we regulate that those received data will be encrypted by disk encryption software if enabled.
As you can see, it is not a pure linear relation between data size and mobile device's energy consumption.
In detail, the energy consumption of the mobile device enabling dm-crypt is close to the mobile device enabling \textit{Populus} when data size is small and gradually changed to linear relation as data size grows larger. 
Due to file system buffer and disk I/O buffer, part of received data may be lazily cached in buffer so that disk encryption may not be fully triggered.
On average, 56\% of energy consumption is cost by dm-crypt and \textit{Populus} can reduce it to 25\%.
\subsection{Evaluation on Pure Disk Encryption/Decryption Operations}
To compare dm-crypt with \textit{Populus}, one effective way is to compute the energy consumption of pure disk encryption operations in dm-crypt and \textit{Populus} and then compute the improvement percentage.
However, both of them can not be directly measured by Monsoon power monitor.
To solve this problem, we design a comparison model to estimate them.

Next, we formally introduce our comparison model.
The energy cost of dm-crypt is denoted by $AE_{i}$ and the energy cost of \textit{Populus} is denoted by $PE_{i}$.
Here, $i$ denotes the file size in certain experiment.
For example, when recording a video, $i$ denotes the video file size.
Let $GE_{i}=\frac{AE_{i}-PE_{i}}{AE_{i}}$ denote the percentage of energy that \textit{Populus} saves in comparison with dm-crypt when processing $i$-megabyte file, and we use $\overline{GE}$, the average of all $GE_{i}$, to compare the energy consumption between \textit{Populus} and dm-crypt.
We regulate three different configurations as: $Conf.1$, all disk encryption systems are disabled; $Conf.2$, only dm-crypt is enabled; $Conf.3$, only \textit{Populus} is enabled.

We first measure the energy consumption $EC_{i,j}$ ($j\in\{1,\ldots,3\}$) and the time cost $ET_{i,j}$ ($j\in\{1,\ldots,3\}$) of our mobile phone with different $conf.j$ ($j\in\{1,\ldots,3\}$).
In addition, we observe that the energy consumption of Android OS is stable, so we denote $SP$ as the energy cost of system per second, and $SP$ can be directly computed by measuring the power consumption when our mobile device is idle.
Then we compute $GE_{i}$ based on $EC_{i,j},j\in\{1,\ldots,3\}$, $ET_{i,j},j\in\{1,\ldots,3\}$, and $SP$.
Let $FE_{i}$ denote the energy consumption of the pure file and disk operations on $i$-byte file excluding disk encryption/decryption and $SE_{i,j},j\in\{1,\ldots,3\}$ denote the energy consumption of Android OS for $Conf.j$.
Considering $SE_{i,j}=SP\cdot ET_{i,j}$,$EC_{i,1}=FE_{i}+SE_{i,1}$, $EC_{i,2}=FE_{i}+SE_{i,2}+AE_{i}$, and $EC_{i,3}=FE_{i}+SE_{i,3}+PE_{i}$, we can compute $GE_{i}$ as follows:
\begin{align}
& GE_{i}=\frac{AE_{i}-PE_{i}}{AE_{i}} \cr
& ~~~~~=\frac{(EC_{i,2}-FE_{i}-SE_{i,2})-(EC_{i,3}-FE_{i}-SE_{i,3})}{EC_{i,2}-FE_{i}-SE_{i,2}} \cr
&~~~~~=\frac{(EC_{i,2}-SE_{i,2})-(EC_{i,3}-SE_{i,3})}{EC_{i,2}-(EC_{i,1}-SE_{i,1})-SE_{i,2}} \cr
&~~~~~=\frac{(EC_{i,2}-EC_{i,3})-(SE_{i,2}-SE_{i,3})}{(EC_{i,2}-EC_{i,1})-(SE_{i,2}-SE_{i,1})} \cr
&~~~~~=\frac{(EC_{i,2}-EC_{i,3})-SP(ET_{i,2}-ET_{i,3})}{(EC_{i,2}-EC_{i,1})-SP(ET_{i,2}-ET_{i,1})} \cr
\end{align}
Then we can compute $\overline{GE}$ with all $GE_{i}$.

To prepare for this experiment, we implement a test APP using JNI technique to invoke random file reading and writing without caching data into various buffer mechanism.
We also turn off irrelevant APPs and sensors and then run our test APP while measuring energy consumption.

\begin{figure}
	\centering
	\includegraphics[height=2in, width=3.4in]{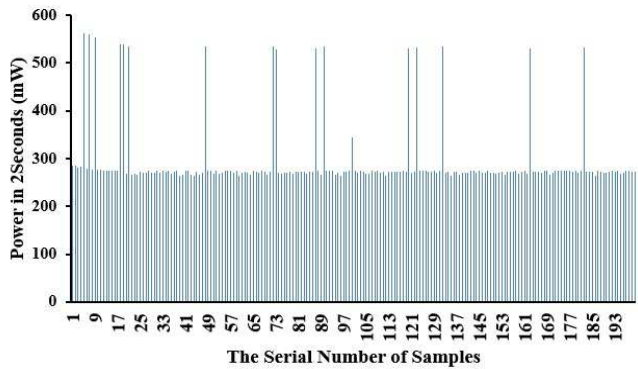}
	\caption{Samples of SP}
	\label{fig:SK}
\end{figure}

\begin{figure}
\centering
\includegraphics[height=2.2in, width=3.4in]{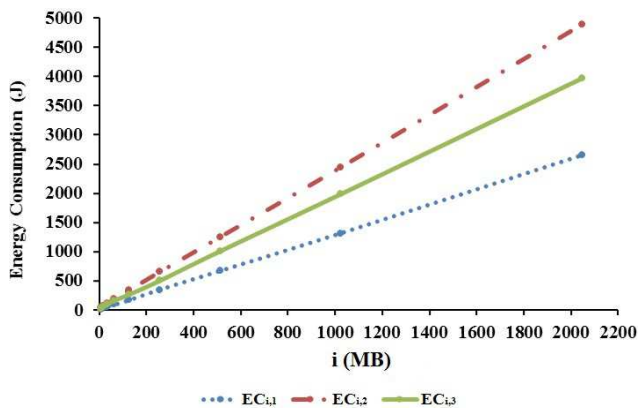}
\caption{Energy consumption of the whole mobile device}
\label{fig:ECC}
\end{figure}

Then we use Monsoon power monitor to observe $SP$, $EC_{i,j}$, $ET_{i,j}$. 
Samples of $SP$ are shown in Fig.~\ref{fig:SK}.
As you can see, most of samples are closed and a few samples are higher than others.
We think those exceptional samples are mainly caused by periodic system scheduling and it doesn't affect our assumption that $SP$ is nearly fixed.
Finally, we average all samples and get 294 milliwatt as the estimation of $SP$.
Fig.~\ref{fig:ECC} shows the observations of $EC_{i,j}$.
The curve shows linear feature of energy consumption.

Based on comparison model, we compute $\overline{GE}$ to show the improvement percentage of energy consumption in \textit{Populus} compared to dm-crypt.
\begin{figure}
\centering
\includegraphics[height=2in, width=3in]{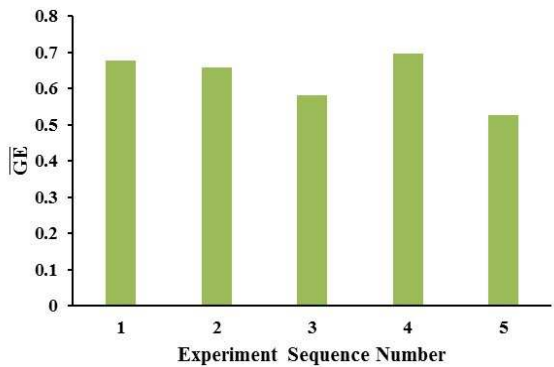}
\caption{Percentage of improvement}
\label{fig:PI}
\end{figure}
Fig.~\ref{fig:PI} shows five $\overline{GE}$ in five repeated experiments.
We can see that $\overline{GE}$ is roughly between 50\% and 70\%.
Therefore, we can conclude that \textit{Populus} saves 50\%-70\% less energy than dm-crypt.

\section{Related Work}\label{sec: related work}
Popular and secure disk encryption software includes dm-crypt (for Linux and Android), BitLocker (for Windows), FileVault (for Mac OS X) and TrueCrypt (for Windows and Linux)~\cite{DiskEncryptionSoftware}.
They conduct encryption/decryption with tweakable scheme~\cite{Shai:Halevi2003A} and SPN-based block ciphers~\cite{shannon1949communication}.
However, we found that tweakable scheme and SPN essentially lead to the energy overhead in disk encryption software and explained it in Section~\ref{sec: background}.
As an attempt to improve efficiency, Crowley and Paul proposed Mercy, a lightweight disk encryption software~\cite{crowley2001mercy}.
Unfortunately, Fluhrer proved that Mercy is insecure in cryptography~\cite{fluhrer2002cryptanalysis}.
\section{conclusion}
In this paper, we develop a kernel-level disk encryption software \textit{Populus}
to reduce the high energy consumption of disk encryption,
which is critical for mobile devices.
We observe that at most 98\% of \textit{Populus's} encryption/dycryption computation is input-free, which can be accomplished in advance during initialization,
so \textit{Populus} is energy-efficient for processing real-time encryption/decryption requests.
We conduct cryptanalysis on \textit{Populus} and find it is computationally secure when facing state-of-the-art cryptanalysis techniques.
We also conduct energy consumption experiments and our experimental results show that \textit{Populus} consumes 50\%-70\% less energy in comparison with dm-crypt.
\bibliographystyle{IEEEtran}
\bibliography{huyang}
\appendices
\section{Cryptanalysis in Linear-Based Algebra Attack}
\label{APPENDEX: Algebra}
In this section, we conduct cryptanalysis to assess whether \textit{Populus} can defend against linear-based algebra attack.
To keep pace with aforementioned sections, we continue to use notations of our security definitions in Section~\ref{sec: sd}.

Linear-based algebra attack usually breaks linear-based ciphers by solving certain linear equation.
For example, $C_{1}=LP_{1}$,$\ldots$,$C_{n}=LP_{n}$ where $L$ is $n\times n$ matrix and $P_{i},C_{i}$ are $n\times 1$ matrices or vectors.
Then attackers can solve $L$ by computing $L=[C_{1},\ldots,C_{n}][P_{1},\ldots,P_{n}]^{-1}$.
In the same way, linear-based algebra attack can break \textit{Populus} if it \textit{collects} $64$ (plaintext,ciphertext) pairs sharing same sector key in $M_{1:\theta}$.
Here, 'collect' denotes that attackers can conduct \textit{chosen-plaintext attack}~\cite{ChosenPlaintextAttack} to get several (plaintext,ciphertext) pairs.

In detail, we assume that attackers can get $r$ pairs $(P^{''}_{1},C^{''}_{1})$, $\ldots$, $(P^{''}_{r},C^{''}_{r})$ where $C^{''}_{i}=E(P^{''}_{i},M^{''}_{i})$.
We define $Event_{\mu},\mu\in\{1,\ldots,\theta\}$ as an event that $M_{\mu}=M^{''}_{i_{1}}=\ldots=M^{''}_{i_{64}}$ where $\mu \in\{1,\ldots,\theta\}$ and $i_{1},\ldots,i_{64}\in\{1,\ldots,r\}$ and $r\geq 64$ and $i_{1},\ldots,i_{64}$ are all different from each other.
If $Event_{\mu}$ happens,
attackers can solve $M_{\mu}$ by computing
\begin{equation}
M_{\mu}=[C{''}_{i_{1}},\ldots,C{''}_{i_{64}}][P^{''}_{i_{1}},\ldots,P^{''}_{i_{64}}]^{-1}
\end{equation}
and then construct the distinguisher as
\begin{eqnarray}
Adv(X_{1:\theta})=
\begin{cases}
0, &D(X_{\mu},M_{\mu})=P_{\mu}\cr
1, &D(X_{\mu},M_{\mu})=P^{'}_{\mu}
\end{cases}
\end{eqnarray}
where $X_{1:\theta}=(X_{1},\ldots,X_{\theta})\subset \{sequences~of~512~bytes\}^{\theta}$.

However, finding \textit{fitted} $\mu,i_{1},\ldots,i_{64}$ (i.e., $\exists \mu (Event_{\mu})$) is either complicate (i.e, with high computational complexity) or hopeless (i.e., with low success probability).
Then we give two lemma and one theorem to prove our statement.
\begin{LM}
For every $M_{1:\theta},P_{1:\theta},P^{'}_{1:\theta}$, we have
\begin{equation}
\mathcal{P}(Event_{\mu})=1-\sum_{l=0}^{63}(_{i}^{r})(\frac{1}{2^{128}})^{l}(1-\frac{1}{2^{128}})^{r-l},
\end{equation}
where $(_{i}^{r})$ denotes the combinatorial number of $i$-combinations in $\{1,\ldots,r\}$.
\begin{IEEEproof}
Let $\alpha(l),l\in\{0,\ldots,r\}$ denote the proposition: $\exists i_{1},\ldots,i_{r}(i_{1},\ldots ,i_{r}~are~different\land M^{''}_{i_{1}}= \ldots=M^{''}_{i_{l}}=M_{0} \land M_{i_{l+1}},\ldots,M_{i_{r}}\neq M_{0})$ where $i_{1},\ldots,i_{r}\in\{1,\ldots,\theta\}$.
Given all $\alpha(l)$ are mutually exclusive from each other, we have
\begin{equation}
\mathcal{P}(Event_{\mu})=1-\sum_{l=0}^{63}(_{l}^{r})\mathcal{P}(\alpha(l)|\mu)
\end{equation}
	
Next, we compute $\mathcal{P}({\alpha(j)}|\mu)$.
From production of sector key, we can get that $\mathcal{P}(M^{''}_{i}=M_{\mu}|\mu)=\frac{1}{2^{128}}$.
For $M^{''}_{i}=M_{0}$ is conditionally independent from $M^{''}_{j}=M_{0}$, $\alpha(l)|\mu$ obeys binomial distribution so that:
\begin{equation}
\mathcal{P}({\alpha(l)}|\mu)=(_{l}^{r})(\frac{1}{2^{128}})^{l}(1-\frac{1}{2^{128}})^{r-l}.
\end{equation}
\end{IEEEproof}
\label{LM: Event1}
\end{LM}
From Lemma~\ref{LM: Event1}, we can use \textit{inclusion-exclusion principle}~\cite{InclusionExclusion} and \textit{Chernoff bound}~\cite{arratia1989tutorial} to infer that
\begin{LM}
For every $M_{1:\theta},P_{1:\theta},P^{'}_{1:\theta}$, we have
\begin{equation}
\mathcal{P}(\exists \mu (Event_{\mu}))\leq \theta e^{-r\mathcal{T}(\frac{64}{r},\frac{1}{2^{128}})},
\end{equation}
where $\mathcal{T}(x,y)=xlog(\frac{x}{y})+(1-x)log(\frac{1-x}{1-y})$ and $r\leq 2^{120}$.
\begin{IEEEproof}
Based on inclusion-exclusion principle~\cite{InclusionExclusion} and Chernoff bound~\cite{arratia1989tutorial}, we have
\begin{align}
\notag
& \mathcal{P}(\exists \mu (Event_{\mu}))\leq \sum_{\mu=1}^{\theta}\mathcal{P}(Event_{\mu}) \cr
& ~~~~~~~~~~~~~~~~~~~~= \theta(1-\sum_{i=0}^{63}(_{i}^{r})(\frac{1}{2^{128}})^{i}(1-\frac{1}{2^{128}})^{r-i}) \cr
& ~~~~~~~~~~~~~~~~~~~~< \theta e^{-r\mathcal{T}(\frac{64}{r},\frac{1}{2^{128}})}
\end{align}	
\end{IEEEproof}
\label{LM: Event12}
\end{LM}
From Lemma~\ref{LM: Event12}, we can get the folloing theorem.
\begin{THM}
\textit{Populus} is $(t,\theta e^{-t\mathcal{T}(\frac{64}{t},\frac{1}{2^{128}})},\theta)$ message indistinguishable from linear-based algebra attack if $t\leq 2^{80}$.
\begin{IEEEproof}
Assume that the distinguisher's computational complexity is not more than $t$.
Then $r\leq t$ because choosing (plaintext,ciphertext) pairs belongs to attackers' computation.
So $r<2^{80}<2^{120}$ when $t<2^{80}$.
Therefore, Lemma~\ref{LM: Event12} can derive that the possibility of a successful distinguish is not more than $\theta e^{-t\mathcal{T}(\frac{64}{t},\frac{1}{2^{128}})}$.
\end{IEEEproof}
\label{THM: MI}
\end{THM}
Supported by the scientific computational software \textit{Wolfram Mathematica}, we get $\theta e^{-t\mathcal{T}(\frac{64}{t},\frac{1}{2^{128}})}\ll \frac{1}{2^{80}}$ when $t\leq 2^{80}$ and $\theta \leq t$.
Hence, Theorem~\ref{THM: MI} implies that linear-based algebra attack can't break \textit{Populus} in reasonable computational complexity.
\end{document}